\documentclass[aps,preprint,nofootinbib]{revtex4}
\usepackage{amssymb}
\usepackage{graphicx}
\usepackage{epsfig}

\def\beq{\begin{eqnarray}}
\def\eeq{\end{eqnarray}}
\def\non{\nonumber}
\def\vep{\varepsilon}
\def\ren{renormalization~}

\begin{document}

\title{ A solution to Higgs naturalness }

\author{
  Zheng-Tao Wei\footnote{Email: weizt@nankai.edu.cn},
  Li-Gong Bian\footnote{Email: ligongbian@gmail.com}}

\affiliation{ School of Physics, Nankai University, Tianjin 300071, China }

\begin{abstract}

The Standard Model (SM) is usually considered to be unnatural because 
the scalar Higgs mass receives a quadratic divergent correction. We 
suggest a new way to solve the naturalness problem from point of view 
of renormalization group method. Our approach is illustrated through 
the familiar $\phi^4$ theory. A renormalization group equation for scalar 
field mass is proposed by introducing a subtraction scale.  We give a 
non-trivial prediction: the Higss mass at short-distance is a damping 
exponential function of the energy scale. It follows from a characteristic 
of the SM that the couplings to Higgs are proportional to field masses, 
in particular the Higgs self-interactions. In the ultraviolent limit, 
the Higgs mass approaches to a mass called by Veltman mass which is at 
the order of the electroweak scale. The fine-tuning is not necessary. 
The Higgs naturalness problem is solved by radiative corrections themselves.

\end{abstract}

\maketitle

\section{Introduction}

The Higgs field plays a fundamental role in the SM. It provides the
origin of spontaneous symmetry breaking and masses of all matter
fields. The crucial purpose of the running Large Hadron Collider
(LHC) is to test this mechanism. However, it was known for a long
time that the scalar field suffers from a problem caused by
quadratic divergence \cite{Wilson:1970ag,Susskind:1978ms}. In particular,
the one-loop correction to the Higgs mass square, is proportional to a large
momentum cut-off $\Lambda^2$ by \cite{Veltman:1980mj,Djouadi:2005gi}
 \beq \label{eq:mH}
 m_H^2=(m_H^0)^2+\frac{3}{8\pi^2 v^2}
  \left [ m_H^2+2m_W^2+m_Z^2-4m_t^2 \right ]\Lambda^2~.
 \eeq
where $m_i$ are masses of gauge and fermion fields.  In order to
satisfies the experimental constraints on the Higgs mass which is
at the order of 100 GeV, a delicate cancelation between the bare mass square
and the counter-term requires an incredible fine-tuning of parameters.
Because there is no symmetry protecting the small Higgs mass in the SM, the Higgs is
considered as "unnatural".

Many proposals have been proposed to solve the naturalness problem
\cite{Giudice:2008bi}. Veltman pointed out a relation
\cite{Veltman:1980mj}
 \beq \label{eq:Ve}
 m_H^2=4m_t^2-2m_W^2-m_Z^2~.
 \eeq
If the above condition holds, the quadratic divergence cancels.
Taking into account of higher order corrections, the Veltman
condition is no longer valid \cite{Djouadi:2005gi}. Supersymmetry
has an attractive property to solve the naturalness problem by
cancelation between fermions and bosons \cite{Witten:1981nf}. But
this symmetry is broken in reality, and the Higgs mass depends on
the supersymmetry breaking scale quadratically. In
\cite{Chaichian:1995ef}, the authors suggest that the cancelation of
quadratic divergence at scale of new physics.

The result given in Eq. (\ref{eq:mH}) contains only the one-loop
contribution. Although seemingly quadratically divergent, it is not
by all means that the final result with all-order radiative
corrections are divergent.  In quantum field theory, there are some
unexpected or non-trivial examples which contradicts the simple
intuition. One classical example is the asymptotic freedom. The
coupling constant of a non-Abelian gauge field theory, e.g. QCD, is
logarithmic divergent in one-loop, while it vanishes in the
short-distance limit. Another example is the elastic form factor of
a fermion at large momentum transfer which is usually called by
Sudakov form factor \cite{Sudakov,Wei:2004uh}. The one-loop
correction contains a large double-logarithm. Summing
double-logarithm to all orders in the coupling constant produces a
rapid damping exponential function. The success of the
two examples relies on renormalization group method. Since the SM is
renormalizable, the Higgs mass is independent of the cut-off scale,
and Eq. (\ref{eq:mH}) does not provide the true scale dependence. 
Moreover, whether the bare mass is really divergent or not is unknown. 
To answer these questions, it is necessary to study the renormalization 
evolution and the short-distance behavior of the Higgs mass.

The dimensional regularization is the most popular method to
regulate the divergence, while the quadratic divergence is absent in
this method because of a definition that scaleless integral is zero.
According to this point, some theorists have the opinion that there
is no naturalness problem at all. In \cite{Veltman:1980mj}, Veltman
pointed out that dimensional regularization is not physical because
theory with space-time dimension $d\neq 4$ is unphysical. We provide
another comment based on Wilson's renormalization 
group method \cite{Wilson:1974mb}. The dimensionless integral $\int
\frac{d^4k}{k^2}$, which is quadratic divergent by dimensional
analysis, sums up the virtual particle contributions with momentum
square from 0 to $\infty$. The larger the momentum square is, the
more important it contributes. If we simply defined this integral to
be zero, the real physics from different energy scales will be
missed. The conventional \ren group equations are usually given in
dimensional regularization where the mass renormalization is
multiplicative. Because of the quadratic divergence, the
renormziation of mass is additive. We have to search for new types of
renormalization group equation.

Nevertheless, a consistent renormalization program for the quadratic
divergence, such as the regularization method and renormalization
scheme, is not mature. Veltamn uses the dimensional regularization,
but chooses the dimension $d$ close to 2 rather than 4 in the
integral because the pole occurs at $d=2$. Then he defines the pole
to be proportional to the momentum cut-off square
\cite{Veltman:1980mj}. This treatment seems to be a combination of
dimensional regularization and momentum cut-off. Another obstacle
concerns the renormalization scheme, i.e. the choice of
renormalization condition. Fujikawa proposed a speculative scheme by
introducing a subtraction scale \cite{Fujikawa}. This scheme is simple 
and has an advantage in deriving the \ren group equations. We will 
discuss this method and use it to discuss the evolution of scalar field 
mass. The implications to the Higgs naturalness is addressed.

\section{ Renormalization and renormalization group equation of $\phi^4$ theory}

The familiar scalar $\phi^4$ theory is simple, and only one field is 
involved. Thus it provides an ideal laboratory to study the renormalization group. 
The unrenormalized Lagrangian is
 \beq
 {\cal L}_0=\frac{1}{2}\left[ \partial_\mu\phi_0\partial^\mu\phi_0
  -m_0^2\phi_0^2 \right]-\frac{\lambda_0}{4!}\phi_0^4~,
 \eeq
we don't discuss the case with spontaneously symmetry breaking, thus
the mass square is positive $m_0^2>0$. The $\phi^4$ theory in four
space-time dimension is renormalizable and the divergences can be
absorbed into the redefinition of the fields and coupling
parameters. The standard renormalization program is to express the
bare quantities in terms of the renormalized ones by
 \beq
 &&\phi_0 = Z_\phi^{1/2}\phi, \qquad  \lambda_0 = Z_\lambda \lambda~, \non\\
 &&m_0^2  = m^2-\delta m^2~.
 \eeq
The filed $\phi$ and the dimensionless coupling constant $\lambda$
are multiplicative renormalized, while the \ren of mass is different: it is 
additive. We can't write $m_0^2=Z_m m^2$ in a conventionally way. The reason
is that the mass correction is quadratic divergent and others are
only logarithmically divergent.

At one-loop order, the self-energy correction is given by
 \beq \label{eq:self}
 -i\Sigma(p^2)=\frac{1}{2}(-i\lambda)\int\frac{d^4 k}{(2\pi)^4}\frac{i}{k^2-m^2+i\vep}~.
 \eeq
The s-channel vertex correction as
 \beq \label{eq:vertex}
 &&\Gamma(p^2)=\Gamma(s)=\frac{1}{2}(-i\lambda)^2\int\frac{d^4 k}{(2\pi)^4}
 \frac{i}{\left[(k-p)^2-m^2+i\vep\right]}\frac{i}{\left(k^2-m^2+i\vep\right)}~,
 \eeq
The notations can be found in textbook \cite{Cheng:2000ct}. We omit
them to simplify the illustration.

In order to see how the quadratic divergences in the self-energy corrections are produced,
let us consider only the integral of the $\Sigma(p^2)$,
 \beq \label{eq:qu}
 \int\frac{d^4 k}{(2\pi)^4}\frac{1}{k^2-m^2+i\vep}=\int\frac{d^4 k}
  {(2\pi)^4}\frac{m^2}{k^2(k^2-m^2+i\vep)}+\int\frac{d^4 k}{(2\pi)^4}\frac{1}{k^2}~.
 \eeq
In the above equation, the first part is logarithmically divergent
and it is proportional to $m^2$. The second part is quadratic
divergent and independent of $m^2$.

We apply the Pauli-Villas regularization to make the integral
finite. For the self-energy correction of Eq. (\ref{eq:self}), the
propagator is modified to
 \beq
 \frac{1}{[k^2-m^2+i\epsilon]}~~ \to ~~\frac{\Lambda^4}{[k^2-m^2+i\vep]
 \left(k^2-\Lambda^2+i\vep\right)^2}~.
 \eeq
where $\Lambda\gg m$ is a large mass parameter. The divergent parts
at zero external momentum $p^2=0$ are
 \beq
 \Sigma(0)=\frac{\lambda}{32\pi^2}\left[\Lambda^2-m^2{\rm ln}\frac{\Lambda^2}{m^2}\right]~,
 \qquad
 \Gamma(0)=i\frac{\lambda^2}{32\pi^2}{\rm ln}\frac{\Lambda^2}{m^2}~.
 \eeq

The basic idea of Fujikawa's renormalization  scheme \cite{Fujikawa}
can be demonstrated by introducing a subtraction scale $\mu$ by the 
simple relations below,
 \beq \label{eq:mu}
 \Lambda^2~ &=& (\Lambda^2-\mu^2)~+~~\mu^2~,\non\\
 {\rm ln}\frac{\Lambda^2}{m^2} &=& ~~~{\rm ln}\frac{\Lambda^2}{\mu^2}~~~~+~
 {\rm ln}\frac{\mu^2}{m^2}~.
 \eeq
The above relations are not just equalities. They represent that the low
energy physics is separated from the high energy part. The introduction of 
scale $\mu$ can be inferred in the dimensional regularization, for instance,
 \beq
 \lambda ~d^4 k\Longrightarrow \lambda~\mu^{4-d}d^d k~.
 \eeq
If the space-time dimension $d$ is 2 as done in
\cite{Veltman:1980mj}, we need a $\mu^2$ associated with $\lambda$
in order to make the coupling constant dimensionless. Thus, the
scale $\mu$ acts a similar role as the renormalization scale in the
dimensional regularization.

As the minimal subtraction in dimensional regularization, our scheme
is also mass-independent. This provides great advantage in
deriving the \ren group equations. Thus, the one-loop results
for \ren constants are
 \beq \label{eq:Z}
 &&Z_\phi = 1+{\cal O}(\lambda^2), \qquad
 Z_\lambda = 1+\frac{3\lambda}{32\pi^2}{\rm ln}\frac{\Lambda^2}{\mu^2}~,\non\\
 &&\delta m^2 = \frac{\lambda}{32\pi^2}\left(\Lambda^2-\mu^2-m^2{\rm ln}
 \frac{\Lambda^2}{\mu^2}\right)~.
 \eeq

The choice of scale $\mu$ is arbitrary and this arbitrariness naturally leads to the
\ren group equations. The unrenormalized field $\phi_0$ and coupling constant $\lambda_0$
are independent of $\mu$, thus
 \beq
 \frac{d\phi_0}{d{\rm ln}\mu}=0, \qquad \qquad
 \frac{d\lambda_0}{d{\rm ln}\mu}=0~,
 \eeq
Two functions can be defined by
 \beq \label{eq:rge}
 \gamma_\phi(\lambda)= \frac{1}{2}\frac{1}{Z_\phi}\frac{dZ_\phi}{d{\rm ln}\mu}~,
 \qquad
 \beta(\lambda)\equiv\frac{d\lambda}{d{\rm ln}\mu}=
    -\lambda\frac{1}{Z_\lambda}\frac{d Z_\lambda}{d{\rm ln}\mu}~.
 \eeq

The \ren group equation for mass should be different because the mass 
\ren is additive. We don't differentiate the renormalized mass $m^2$ with 
respect to ${\rm ln}\mu$ but with $\mu^2$. The bare mass is independent of
$\mu$, thus
 \beq
  \mu^2\frac{dm_0^2}{d\mu^2}=0, \qquad \to~~~~ \mu^2\frac{dm^2}{d\mu^2}=
  \mu^2\frac{d(\delta m^2)}{d\mu^2}~,
 \eeq
From Eq. (\ref{eq:Z}), $\delta m^2$ contains both $\mu^2$ and $m^2$
terms which correspond to quadratic and logarithmic divergences,
respectively. Thus, we define two \ren group functions $\gamma_\mu$
and $\gamma_m$ by
 \beq \label{eq:m}
 \mu^2\frac{dm^2}{d\mu^2}=\gamma_\mu(\lambda)\mu^2-\gamma_m(\lambda) m^2~,
 \eeq
A minus sign is added in the $\gamma_m$ term in order to accord with
the conventional definition (differs by a factor of 2). In the adopted \ren 
scheme, the functions $\gamma_\mu$ and $\gamma_m$ are not explicit
$\mu$-dependent but are functions of $\lambda(\mu)$.  Each of them can be
interpreted as anomalous dimension of mass. $\gamma_\mu$ represents
anomalous dimension induced by the quadratic divergence.

From our calculations, the \ren group functions are obtained to be
 \beq
 \gamma_\mu=-\frac{\lambda}{32\pi^2}~, \qquad \qquad
 \gamma_m=-\frac{\lambda}{32\pi^2}~.
 \eeq
To solve the Eq. (\ref{eq:m}) is difficult, we restrict our
discussions at short-distance where $\mu^2\gg m^2$ and only the
$\gamma_\mu$ term is retained. But, even that, we still cannot give
an analytic solution because the scale dependence of $\lambda(\mu)$.
Note that the linear dependence of $\mu^2$ is more important than
the logarithmic dependence when $\mu$ is large, it is reasonable to
neglect the variation of $\lambda$ with $\mu$. Under this
approxiamtion, we obtain
 \beq
 m^2(\mu)=m^2(\mu_0)-\frac{\lambda}{32\pi^2}\left(\mu^2-\mu_0^2\right)~,
 \eeq
The renormalized mass square $m(\mu)^2$ is a linear function of $\mu^2$.
When $\mu$ increases, $m^2$ decreases. This decreasing is ascribed
to the negative sign of $\gamma_\mu$. There exists a possibility
that the induced mass square becomes negative when $\mu$ is large. This
case is related to spontaneous symmetry breaking and we left it for
a future study.

The renormalization group equation can also be derived from another
way through the $\Lambda$-dependence. The fact that renormalized
mass is independent of $\Lambda$ leads to
 \beq \label{eq:Lambda}
 \Lambda^2\frac{dm_0^2}{d\Lambda^2}=\gamma_\mu(\lambda)\Lambda^2-\gamma_m(\lambda) m^2~.
 \eeq
The understanding of renormalization group equation in this way is
not new and have implied in the textbook of Zee \cite{Zee:2003mt}.
Because the renormalized and bare masses satisfies the same evolution 
equation, we deduce an interpretation: the bare mass is 
nothing but the renormalized mass by taking the scale $\mu$ 
to $\Lambda$ (or $\infty$) in a cut-off regularization. In other words, 
the bare mass contains virtual particle momentum from 0 to $\Lambda$ and the 
renormalized mass contains momentum from 0 to $\mu$. Because the
experiment has limited resolution, it seems that the renormalized mass is 
more important. The above interpretation applies for any bare quantities, i.e. 
the bare quantities are the renormalized ones at the ultraviolent limit.

The \ren group equation for Green function can be obtained
straightforwardly. Denote $G_{n}^0(p,\lambda_0,m_0)$ and
$G_{n}(p,\lambda,m,\mu)$ by the unrenormalized and renormalized
truncated (amputated) connected n-point Green function,
respectively. Multiplicative \ren of field $\phi$ gives
$G_n^0=Z_\phi^{-n/2}G_n$. The unrenormalized Green function does not
depend on $\mu$, thus $\mu\frac{d}{d\mu}G_n^0=0$. The \ren group
equation for $G_{n}(p,\lambda,m,\mu)$ is
 \beq
 \left[\mu\frac{\partial}{\partial\mu}+\beta\frac{\partial}{\partial \lambda}
  +2\left(\gamma_\mu\mu^2-\gamma_m m^2\right)\frac{\partial}
  {\partial m^2}-n\gamma_\phi\right]G_n(p,\lambda,m,\mu)=0~.
 \eeq
The \ren group functions $\beta,~ \gamma,~ \gamma_\mu,~ \gamma_m$
have been defined in Eqs. (\ref{eq:rge}) and (\ref{eq:m}). The
solution of the above equation is similar to the conventional one except 
that the renormalized mass satisfies the new evolution equation.

\section{The evolution of the Higgs mass}

Now, we turn to the Higgs mass. According to Eq. (\ref{eq:mH}) and
our treatment of quadratic divergence, the one-loop correction gives
 \beq
 \delta m_H^2=\frac{3}{8\pi^2 v^2}
  \left [ m_H^2+2m_W^2+m_Z^2-4m_t^2 \right ](\Lambda^2-\mu^2)~,
 \eeq
The renormalization group equation for the renormzlied Higgs mass is
obtained by differentiate the above equation with respect to
$\mu^2$, thus
 \beq
 \frac{dm_H^2}{d\mu^2}=\gamma_\mu^H~,
 \eeq
where the mass anomalous dimension $\gamma_\mu^H$ is given by
 \beq
 \gamma_\mu^H=-\frac{3}{8\pi^2 v^2}
  \left [ m_H^2+2m_W^2+m_Z^2-4m_t^2 \right ]~.
 \eeq
In the anomalous dimension $\gamma_\mu^H$, the boson field
contribution is negative, while the fermion field part is positive.
Compared to the pure scalar field theory, the anomalous dimension
$\gamma_\mu^H$ is proportional to masses of different fields. This
is a special property of the SM where all masses are produced from
the spontaneously symmetry breaking by unsymmetric vacuum and the
fields are coupled to the Higgs proportionally to their masses. Note
that it is just this property which makes the Higgs mass stable.

Let us introduce a mass parameter $m_V$ as
 \beq
 m_V=\sqrt{4m_t^2-2m_W^2-m_Z^2}~.
 \eeq
Here the subscript "V" is borrowed from the name of Veltman, and we
may call $m_V$ by "Veltman mass". If we use the experimental masses,
$m_V\simeq 310$ GeV. Note that the masses appeared in the above
equations are renormalized masses rather than the experimentally
observed masses.

The solution of the Higgs mass is obtained as
 \beq \label{eq:mH2}
 m_H^2(\mu)=m_V^2(\mu)+\left[ m_H^2(\mu_0)-m_V^2(\mu_0) \right]
  {\rm exp}\left\{-\frac{3}{8\pi^2 v^2}(\mu^2-\mu_0^2)\right\}~.
 \eeq
where $\mu_0$ is an initial energy scale. We have neglected $\mu$
dependence of $m_W, m_Z, m_t$ since their dependence is
logarithmical. The solution of the Higgs mass is an exponential
damping function. It falls very fast. When
$\frac{\mu^2-\mu_0^2}{v^2}=8\pi^2\approx 80$, ${\rm
exp}\left\{-\frac{3}{8\pi^2 v^2}(\mu^2-\mu_0^2)\right\}\approx
0.05$, the Higgs mass $m_H^2(\mu)$ is very close to $m_V^2(\mu)$.
Here, a phenomenon analogous to the Sudakov form factor is
reappeared. The exponentiation is because the anomalous dimension
$\gamma_\mu^H$ is proportional to masses of different fields,
especially the Higgs mass.

In the short-distance limit, i.e., $\mu\to \infty$, we have
 \beq
 m_H^2\to m_V^2=4m_t^2-2m_W^2-m_Z^2~.
 \eeq
Compared to Eq. (\ref{eq:Ve}), Veltman's condition is revived not at
the electroweak scale but in the short-distance limit. As we discussed in 
the previous section, the bare mass is the mass in the short-distance 
limit. Thus, the bare Higgs mass $m_H^0=m_V(\mu=\infty)$  
is at the order of the electroweak scale within perturbation theory. 
There is no quadratic divergence. The Higgs mass at low energy does not 
receive quadratic divergence and the fine-tuning is not necessary. The 
problem of the Higgs naturalness aroused by one-loop correction is rescued by 
radiative corrections themselves. 

In the above result, we have neglected the logarithmic corrections.
Taking into account them will not modify our conclusion because they
are negligible compared to the quadratic terms in the short-distance
limit.

\section{Conclusion}

In this study, we have explored the Higgs naturalness problem. Our
approach is renormalization group method. Because of the quadratic
divergence, the renormalization of the scalar filed mass is
additive, the conventional multiplicative renormalization is not
applicable. A new type renormalization group equation is required for the
scalar field mass.  Using a subtraction scale, it is possible to study 
evolution of the mass. An anomalous dimension for mass
associated with the quadratic divergence is defined. Then the
established renormalization group approach is applied to the Higgs
mass. We find a surprising and maybe non-trivial result: the Higgs
mass at short-distance is not divergent but an exponential damping
function of energy scale. In the short-distance limit, the Higgs
mass approaches to a finite mass which we call the "Veltman mass".
This mass is at the order of the electroweak scale if the
perturbation theory of the SM is valid. The Higgs bare mass is
finite, and the fine-tuning is not needed. The SM is peculiar because
the couplings are proportional to masses. It is this peculiarity which
makes the Higgs mass at the electroweak scale.

In conclusion, the Higgs mass is protected by radiative corrections.
The SM Higgs is natural.

\section*{Acknowledgments}

We thank Yi Liao for valuable discussions and Center for High
Energy Physics, Peking University for its hospitality in a visit
during which some of this research is done. This work was supported
in part by National Natural Science Foundation of China (NNSFC)
under contract No. 10705015.

{\it  Note added} ~~~~After we put the manuscript of this work on arXiv (1104.2735),
we saw Fujikawa's paper where his renormalization method is given explicitly
\cite{Fujikawa:2011zf}. From \cite{Fujikawa:2011zf} and the references therein, one
can see that there had been many positive attempts to treat renormalization of
quadratic divergences of the $\phi^4$ theory and some formulae are very similar to ours.
However, it should be noted that nearly all of them, except Fujikawa's talk at Nankai
University, have no effects on our research.


\begin{thebibliography}{99}

\bibitem{Wilson:1970ag}
  K.~G.~Wilson,
  Phys.\ Rev.\  D {\bf 3}, 1818 (1971).

\bibitem{Susskind:1978ms}
  L.~Susskind,
  Phys.\ Rev.\  D {\bf 20}, 2619 (1979).

\bibitem{Veltman:1980mj}
  M.~J.~G.~Veltman,
  Acta Phys.\ Polon.\  B {\bf 12}, 437 (1981).

\bibitem{Djouadi:2005gi}
  A.~Djouadi,
  Phys.\ Rept.\  {\bf 457}, 1 (2008)
  [arXiv:hep-ph/0503172].

\bibitem{Giudice:2008bi} For a review, see:
  G.~F.~Giudice,
  arXiv:0801.2562 [hep-ph].

\bibitem{Witten:1981nf}
  E.~Witten,
  Nucl.\ Phys.\  B {\bf 188}, 513 (1981).

\bibitem{Chaichian:1995ef}
  M.~Chaichian, R.~Gonzalez Felipe and K.~Huitu,
  Phys.\ Lett.\  B {\bf 363}, 101 (1995)
  [arXiv:hep-ph/9509223].

\bibitem{Sudakov} V. Sudakov, Zh. Eksp. Teor. Fiz. {\bf 30}, 87
  (1956); Sov. Phys. JETP. {\bf 3}, 65 (1956).

\bibitem{Wei:2004uh}
  Z.~T.~Wei,
  arXiv:hep-ph/0403069.

\bibitem{Wilson:1974mb}
  K.~G.~Wilson,
  Rev.\ Mod.\ Phys.\  {\bf 47}, 773 (1975).

\bibitem{Fujikawa}
  K. Fujikawa, seminar given at Nankai University, 2009.

\bibitem{Cheng:2000ct}
  T.~P.~Cheng and L.~F.~Li,
  ``Gauge theory of elementary particle physics: Problems and solutions,''
{\it  Oxford, UK: Clarendon (2000) 306 p.}

\bibitem{Zee:2003mt}
  A.~Zee,
  ``Quantum field theory in a nutshell,''
{\it  Princeton, UK: Princeton Univ. Pr. (2010) 576 p.}

\bibitem{Fujikawa:2011zf}
  K.~Fujikawa,
  arXiv:1104.3396 [hep-th].




\end{thebibliography}
\end{document}